# Charge Separation at Mixed-Dimensional Single and Multilayer MoS$_2$/Silicon Nanowire Heterojunctions

*Alex Henning[§], Vinod K. Sangwan[§], Hadallia Bergeron[§], Itamar Balla[§], Zhiyuan Sun[§], Mark C. Hersam[§,#,∥], Lincoln J. Lauhon[§,\*]*

[§]Department of Materials Science and Engineering, Northwestern University, Evanston, Illinois 60208, USA

[#]Department of Chemistry, Northwestern University, Evanston, Illinois 60208, USA

[∥]Department of Electrical Engineering and Computer Science, Northwestern University, Evanston, Illinois 60208, USA

[\*]E-mail: lauhon@northwestern.edu

**ABSTRACT** Layered two-dimensional (2-D) semiconductors can be combined with other low-dimensional semiconductors to form non-planar mixed-dimensional van der Waals (vdW) heterojunctions whose charge transport behavior is influenced by the heterojunction geometry, providing a new degree of freedom to engineer device functions. Towards that end, we investigated the photoresponse of Si nanowire/MoS$_2$ heterojunction diodes with scanning photocurrent microscopy and time-resolved photocurrent measurements. Comparison of *n*-Si/MoS$_2$ isotype heterojunctions with *p*-

Si/MoS$_2$ heterojunction diodes under varying biases shows that the depletion region in the *p-n* heterojunction promotes exciton dissociation and carrier collection. We measure an instrument limited response time of 1 μs, which is 10 times faster than previously reported response times for planar Si/MoS$_2$ devices, highlighting the advantages of the 1-D/2-D heterojunction. Finite element simulations of device models provide a detailed understanding of how the electrostatics affect charge transport in nanowire/vdW heterojunctions and inform the design of future vdW heterojunction photodetectors and transistors.

**INTRODUCTION** This paper explores photogenerated charge carrier separation at MoS$_2$/Si nanowire heterojunctions, and uses scanning photocurrent microscopy (SPCM) and finite element modeling to show how the band offsets and MoS$_2$ depletion region influence the photoresponse magnitude and speed. Multilayer *n*-MoS$_2$/*p*-Si nanowire heterojunction devices exhibit an instrumentally-limited photoresponse time ≤ 1 μs, which is an order of magnitude lower than the fastest reported response time for planar Si/MoS$_2$ photodiodes.[1,2] More generally, the performance advantages of *mixed-dimensional heterostructures* can be readily explored by integrating layered van der Waals (vdW) materials with other low-dimensional materials, included 0-D quantum dots and 1-D nanowires, regardless of mismatches in lattice constants.[3] For example, PbS quantum dots enhance the photoresponsivity of vdW materials in 0-D/2-D heterojunctions,[4-6] and 1-D/2-D heterojunctions between *p*-type single-walled carbon nanotubes and *n*-type monolayer (1-L) MoS$_2$ show tunable rectification and anti-ambipolar transfer characteristics due to reduced electrostatic screening in ultrathin *p-n* heterojunctions.[7] Reduced screening in ultrathin devices has also been exploited to create gate-tunable Schottky barriers (barristors) in graphene/silicon 2-D/3-D heterostructure devices, showing that 2-D material integration can augment current Si-based technology.[8] More recent studies have demonstrated that 2-D/3-D *n*-MoS$_2$/*p*-silicon type-II heterojunction diodes function as solar cells[9-11] and photodetectors.[1, 12, 13]

The focus of this work is 1-D/2-D heterojunctions, where 1-D refers to the aspect ratio of the semiconductor nanowire, rather than the dimensionality of the band structure (i.e., 1-D sub-bands are not relevant to the work described here). Several 1-D/2-D heterojunctions have been studied recently, including a ZnO nanowire/black phosphorus type-I heterojunction,[14] a GaN nanowire/MoS$_2$ field-effect transistor,[15] and a ZnO nanowire/WSe$_2$ diode whose photoresponse was characterized at visible and near-infrared wavelengths.[16] However, the role of the depletion region in determining the magnitude and speed of the device response was not examined in detail. Free carrier generation and separation at a heterojunction are influenced by the band discontinuities, which promote exciton dissociation, and by the depletion region adjacent to the heterojunction, which varies with applied bias. Here we exploit type-II heterojunctions to promote the dissociation of excitons, which are relatively strongly bound in 2-D materials.[17, 18] Furthermore, we use the tunability of the depletion region to show that the space-charge region in the 2-D material of the *p-n* heterojunction is responsible for the fast photoresponse. Unlike a purely vertical *p-n* vdW heterojunction in which the depletion region extends perpendicular to the interface, the non-planar nanowire/vdW semiconductor heterojunction generates lateral (in-plane) depletion due to incomplete screening of electric fields in the vdW semiconductor[19] and the presence of unscreened (fringing) electric fields. Finite element device simulations and scanning photocurrent microscopy (SPCM) of nanowire/MoS$_2$ *p-n* and *n-n* devices are used to isolate the roles of the heterojunction and depletion region in photocurrent generation. Compared to recent 2-D/3-D vdW heterojunctions, 1-D/2-D heterojunctions create a larger depletion region that results in enhanced photocurrent and faster photoresponse.

**RESULTS and DISCUSSION** Silicon nanowire/$MoS_2$ heterostructures were fabricated to investigate the influence of electrostatics in non-planar device geometries due to the ease of device fabrication, materials stability, and high absorption of $MoS_2$ at visible wavelengths. Furthermore, *n*-type and *p*-type silicon nanowires can be grown with excellent control over diameter and length.[20] Figure 1a shows a typical device layout in which $MoS_2$ was transferred onto Si nanowires that were deposited on substrates (300 nm $SiO_2$ on Si) prior to electrical contact formation. The electrodes are defined by electron-beam lithography and subsequent metal evaporation (titanium/gold; 3 nm/50 nm). The $MoS_2$ monolayer overlaps the Si nanowire, conforming to its sidewalls to form the heterostructure (Fig. 1b). The height difference of 1 nm between the bare and $MoS_2$ covered surface (Fig. 1c) was measured by atomic force microscopy (AFM) and is typical for monolayer $MoS_2$ deposited on $SiO_2$. Raman spectroscopy of the same area (Fig. S1a) confirms coverage of an intact monolayer as detailed in section 1 of the supporting information. Next, we explore how the atomically thin and conformal $MoS_2$/nanowire heterojunction promotes charge carrier separation.

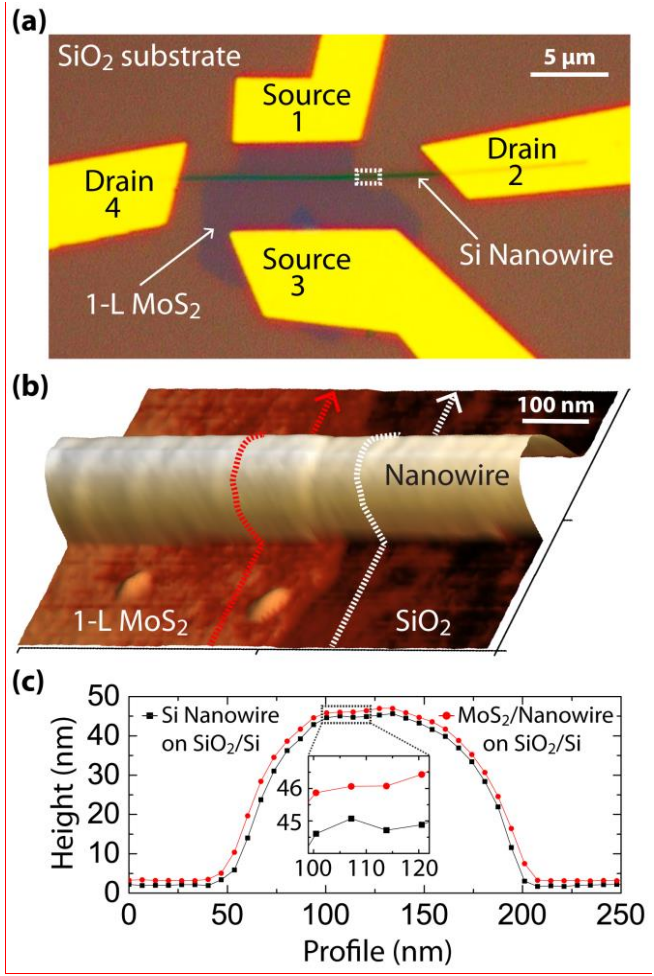

**Figure 1.** (a) Optical micrograph of a device in which a $MoS_2$ monolayer was transferred onto a *p*-Si nanowire followed by contact fabrication. (b) AFM topography of the area indicated by the white rectangle in (a). (c) AFM height profiles across a bare nanowire segment (white arrow in b) and a segment with monolayer $MoS_2$ covering the nanowire surface conformally (red arrow in b).

A boron-doped Si nanowire and conformal MoS$_2$ monolayer form a rectifying *p-n* heterojunction whose current-voltage (*I-V*) characteristics were compared with both *n-n* heterojunction devices and finite element simulations to identify the key factors controlling interfacial charge transport (Fig. 2). The drain current versus drain voltage ($I_D$-$V_D$) of the *p*-Si/MoS$_2$ device (Fig. 2a, red squares) has a rectification ratio of 400 consistent with *p-n* heterojunction formation. The reverse bias saturation current $I_{rev} \leq$ 0.3 nA at $V_D$ = -1 V is in the same range as reported for 1-L MoS$_2$/bulk *p*-Si heterojunctions (< 1 nA).[10, 21] The more symmetrical $I_D$-$V_D$ curve of the *n*-doped nanowire/MoS$_2$ device (Fig. 2a, blue circles) indicates that the rectification of the *p-n* heterojunction device arises from the heterojunction itself, rather than metal-semiconductor interfaces. Additional evidence is provided below.

Finite element simulations of a 2-D device model (Fig. 2b) reproduce key characteristics of the experimental $I_D$-$V_D$ curves for the *p*-Si nanowire/MoS$_2$ (Fig. 2a, solid black line) and the *n*-Si nanowire/MoS$_2$ devices (Fig. 2a, dashed black line). We first describe the simulation details before discussing the comparison. The band alignment in Figure 2c was determined using previously reported electron affinities ($\chi_{Si}$ = 4.07 eV and $\chi_{MoS2}$ = 4.28 eV)[22, 23] and band gaps ($E_{g,Si}$ = 1.12 eV and $E_{g,MoS2}$ = 2.15 eV[24]) as the input parameters for Si and 1-L MoS$_2$, respectively, while noting that the optical band gap of monolayer MoS$_2$ ($E_{exc}$ = 1.88 eV[25]) is lower than the electronic band gap due to the exciton binding energy.[24, 26, 27] Charge transport is modeled with drift-diffusion and takes into account thermionic emission over the energy barrier due to the conduction band offset $\Delta E_C = \chi_{MoS2} - \chi_{Si}$ = 0.21 eV. Monolayer MoS$_2$ is modeled as a 7 Å thin semiconductor with a 3-D density of states[28] (as motivated previously[29]) and a donor-type impurity concentration, $N_D$, of 1.5×10$^{18}$ cm$^{-3}$.[29] The Si nanowire is modeled with a radial doping profile[30] in which the boron (phosphorus) density reaches a maximum of $N_A=1\times10^{19}$ cm$^{-3}$ in the vicinity of the *p*-Si (*n*-Si) nanowire surface.[31] Ohmic contacts were simulated by implementing a Schottky model for the metal/semiconductor interfaces, and tunneling through the Schottky barrier was simulated using the Wentzel–Kramers–Brillouin approximation. Schottky barrier heights, $\Phi_B$, of 0.33 eV,[32] 0.5 eV[33] and 0.6 eV[33] were used for Ti/MoS$_2$, Ti/*n*-Si and Ti/*p*-Si interfaces, respectively.

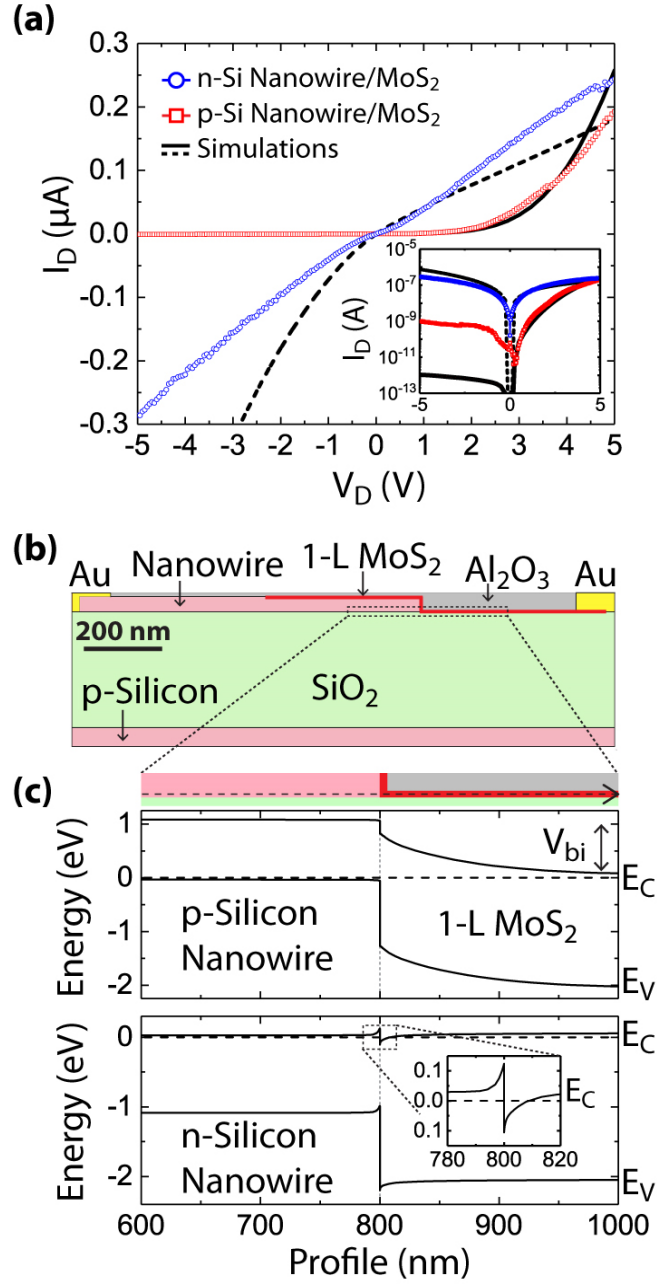

**Figure 2.** (a) Linear and semi-logarithmic (inset) plots of measured (red and blue symbols) and simulated (solid and dashed lines) $I_D$-$V_D$ characteristics while the substrate was grounded. (b) A 2-D device geometry used in finite element simulations. (c) Equilibrium band profiles along the nanowire and overlapping MoS$_2$ (thin red region marked by the dashed arrow in the schematic device geometry) for *p*-Si nanowire/MoS$_2$ (top) and *n*-Si nanowire/MoS$_2$ heterojunctions (bottom).

The band profile in thermal and chemical equilibrium shows the anticipated type-II band alignment and reveals a lateral depletion region of ~200 nm width in the MoS$_2$ layer adjacent to the *p*-Si nanowire (Fig. 2c, top). The simulated built-in potential of 0.8 eV approaches the maximum value, $V_{bi,max} = E_{g,MoS2} - E_{g,Si} - \Delta E_C = 0.82$ eV due to the high dopant concentration in the nanowire, and is responsible for the rectifying characteristics of the heterojunction. Recently, $V_{bi}$ of a bulk Si/1-L MoS$_2$ heterojunction was measured to be ~0.4 eV based on work function differences measured by ultraviolet photoelectron spectroscopy;[9] in that case, the Si wafers were doped at much lower levels ($3.2\times10^{16}$ cm$^{-3}$ < $N_A$ < $2.5\times10^{17}$ cm$^{-3}$)[9] than the nanowires in the present study ($N_A = 1\times10^{19}$ cm$^{-3}$). In comparison, the simulated built-in voltage of the *n*-doped Si nanowire/MoS$_2$ device is very small, and the depletion layer width in the MoS$_2$ is less than 10 nm (Fig. 2c, bottom). The small rectification ratio of the *n*-doped Si nanowire/MoS$_2$ device (Fig. 2a, blue circles) is not surprising given the expected small heterointerface barrier to electron tunneling of $\Delta E_C = 0.21$ eV. We note that the calculated electron affinity of 1-L MoS$_2$ (4.28 eV) may be influenced by the substrate and dielectric adlayer, as well as surface adsorbates (e.g., H$_2$O) that create interfacial dipoles.[29, 34] We also note that a thin native oxide may form between the nanowire and MoS$_2$ in both devices. Considering these effects, the simulation is in reasonable agreement with the measurement. We conclude that the Si nanowire doping establishes the band alignment, which in turn determines the *I-V* characteristics, and one would expect depletion at the Si nanowire/MoS$_2$ heterojunction to produce an internal electric field that promotes charge carrier separation, as described below.

The current was measured as a function of bias under global (full device) illumination and in the dark (Fig. 3a, inset); subtracting the *I-V* curves taken in the light and the dark gives the net photocurrent $I_{ph}$ (Fig. 3a). $I_{ph}$ is larger under reverse ($V_D$ = -2 V) than under forward bias ($V_D$ = 2 V), which is characteristic of a *p-n* heterojunction.[35] In general, the spatial extent of the space-charge region and the magnitude of the electric field increase with increasing reverse bias, which facilitates charge carrier separation and increases the collected drift current. The short-circuit photocurrent, $I_{SC}$, is -55 pA and the open-circuit voltage, $V_{OC}$, is 0.65 V at a gate voltage $V_G$=0 V. The magnitude of $V_{OC}$ is discussed further below. Scanning photocurrent microscopy was used to map the photoactive regions of the device,[36] and analyze the dependencies on the bias ($V_D$) applied across the heterojunction (Figs. 3b and S3), to identify the photocurrent generation regions and mechanisms. SPCM measurements were conducted with a focused laser beam (20 μW/cm$^2$, $\lambda$= 500 nm, full-width at half-maximum, FWHM, of ~0.8 μm) at various biases. The photoresponse at $V_D$= 0 V is dominated by the nanowire/MoS$_2$ heterojunction (Fig. 3b, left), supporting the claim that charge carrier separation occurs due to internal fields at the *p-n* heterojunction. When the *p-n* heterojunction is forward-biased (Fig. 3b, right), which reduces the band bending near the heterojunction (Fig. 3e), the magnitude of local photocurrent is reduced and becomes comparable to that of the reverse-biased source contact (at $V_D$=2 V). Figure 3c quantitatively compares photocurrents at the nanowire/MoS$_2$ heterojunction and MoS$_2$/metal interface for different biases. The positive photocurrent at the MoS$_2$/metal interface (Fig. 3b, right**)** of the grounded source contact is consistent with previous reports[29, 36, 37] of carrier separation by the built-in electric field (photovoltaic effect) at this interface.[29, 36, 38] For completeness, we note that photothermal effects may also contribute to a positive photocurrent at the contacts,[37] but our focus here is on the response of the vdW heterojunction.

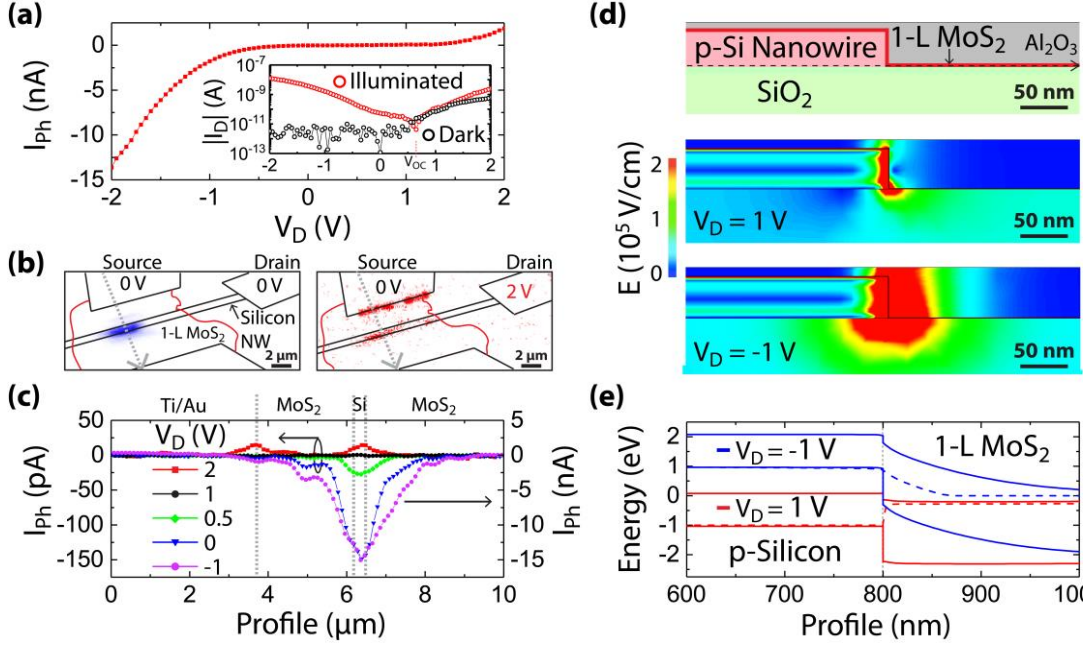

**Figure 3.** (a) Photocurrent as a function of bias under illumination. The inset shows a semi-logarithmic plot of the absolute drain current as a function of the drain voltage in the dark (black) and under illumination (red) with visible light ($\lambda$ = 500 nm, 20 µW/cm$^2$) (b) Photocurrent maps with the drain electrode on nanowire at $V_D$ = 0 V (left panel) and $V_D$ = 2 V (right panel) with the MoS$_2$ source contact at ground in each case. The photocurrent minimum (maximum) in the left (right) panel SPCM image is –180 pA (20 pA). (c) Photocurrent profiles at $V_D$ = 2, 1, 0.5, 0 and -1 V, respectively, taken across the nanowire axis marked by the dashed arrows in (b). (d) Maps of lateral electric fields from 2-D device simulations under biases of $V_D$= 1 V and $V_D$ = -1 V. (e) Simulated band profiles for the two simulations shown in (d).

Photoexcitation at $\lambda$ = 500 nm generates excitons in both Si and MoS$_2$. While thermal excitations at room temperature are sufficient to dissociate the weakly bound excitons in Si, the binding energy of neutral excitons in monolayer MoS$_2$ is higher ($\geq$ 0.2 eV [17, 24, 26, 27]). In our devices, exciton dissociation is expected to occur at the potential discontinuity at the silicon/MoS$_2$ heterojunction, as has been observed previously at vdW heterointerfaces.[39, 40] However, the lateral depletion region in MoS$_2$ also promotes dissociation, which can be mediated by defects at the MoS$_2$/SiO$_2$ interface.[40-42] Indeed, electric fields of kV/cm facilitate exciton dissociation in vdW semiconductor devices,[43] and the dissociation rates at high electric fields >10$^6$ V/cm exceed those of competing mechanisms[44] such as defect-mediated Auger scattering.[41, 42, 45] Consistent with this picture, we will show that the space charge region in the MoS$_2$ of the 1-D/2-D *p-n* heterojunction plays a dominant role in the magnitude and speed of the photoresponse.

The increase in photocurrent magnitude generated by the *p-n* heterojunction in reverse bias is attributed primarily to the evolution of the electric field (space charge region) in the MoS$_2$ adjacent to the heterojunction (Figs. 3c–e), which is influenced by unscreened fringing fields; the field in the heavily doped Si nanowire is less affected by the bias. Minority electrons generated in the Si nanowire may diffuse to the heterojunction and be collected in the MoS$_2$. Some fraction of the excitons generated in the vicinity of the MoS$_2$ depletion region are dissociated at the interface by defects and/or field-assisted processes to produce majority electrons and minority holes; minority holes in MoS$_2$ created within the diffusion length (450 nm[46]) of the depletion region contribute to the observed negative photocurrent at $V_D$ = 0 V as they are swept to the heterojunction, and majority electrons contribute through drift in the opposite direction.[13] The absolute photocurrent generated at the *p*-Si nanowire/MoS$_2$ heterojunction at $V_D$ = -1 V is two orders of magnitude larger than the photocurrent at $V_D$ = 0 V (Fig. 3c) due to the extension of the depletion region in MoS$_2$, which increases the collection area. Modeling shows that the reverse biasing generates fringing electric fields of the order of $10^5$ V/cm that increase the depletion region of MoS$_2$ regions via the field-effect (Fig. 3d), also revealed by the increased spatial extent of the band bending (Fig. 3e, blue curve). Accordingly, the FWHM of the photocurrent profile increases as $V_D$ is decreased from 0 V (Fig. 3c, blue curve) to -1 V (Fig. 3c, purple curve), and the absolute photocurrent generated at and near the *p-n* heterojunction increases by two orders of magnitude due to the extension of the depletion region in MoS$_2$. The photoresponse and bias dependence of *n*-Si nanowire/MoS$_2$ heterojunctions are much weaker due to the absence of a *p-n* heterojunction (Fig. S3).

The short-circuit photocurrent under global (-55 pA) and local (-150 pA) illumination are both negative and of comparable magnitude, indicating that the *p-n* heterojunction dictates the photoresponse (i.e., illumination beyond the depletion region does not contribute significantly). The open-circuit voltage of 0.65 V measured under global illumination (Fig. 3a, inset) is less than the built-in potential of the *p-n* heterojunction due to minority carrier recombination. To estimate the built-in potential, $V_{bi}$, we note that the local photocurrent is zeroed out at $V_D = \sim1$ V (Fig. 3c, black circles), which, based on the interpretation above, should eliminate the depletion region that collects carriers. The simulated band profiles at $V_D = 1$ V are also nearly flat, indicating that $V_{bi}$ corresponds approximately to the flat band potential (i.e., the applied drain voltage at which the internal electric field is effectively screened). The measured $V_{bi}$ of ~1 V (Figs. 3c and S3) is consistent with previous reports for bulk Si/1-L MoS$_2$ devices.[21] However, it is important to note that competing photocurrent generation mechanisms such as the photothermoelectric effect are likely present but not dominating the photocurrent in the investigated *p-n* heterojunction devices. If there is a photocurrent contribution from the photothermoelectric effect, then it would be positive due to the difference in the Seebeck coefficients between the silicon nanowire (~ 200 µV/K [47]) and MoS$_2$ (-4×10$^2$ down to $-1 \times 10^5$ µV/K depending on the back gate voltage[37]). Since the measured photocurrent of the *p*-Si/MoS$_2$ device is negative (for $V_D <$ ~1 V), we conclude that the photovoltaic effect dominates. Therefore, the electric field generated at the *p-n* heterojunction can be exploited for fast photodetection, as we next describe.

Time-resolved photocurrent measurements were carried out on four types of devices: *p*-Si nanowire/*n*-MoS$_2$ monolayer devices, *n*-Si nanowire/*n*-MoS$_2$ monolayer devices, *n*-MoS$_2$ monolayer metal-semiconductor-metal (MSM) photoconductors, and *p*-Si nanowire/*n*-MoS$_2$ multilayer devices. The photocurrent was measured with a low-noise preamplifier with a maximum bandwidth of 50 kHz and one of two high-speed current amplifiers with bandwidths of 500 kHz and 40 MHz (see Materials and Methods Section). Generation, recombination, and transport processes on sub-nanosecond timescales, including carrier drift times, are beyond the resolution of the measurements described here.[48, 49] We first compare the *p-n* heterojunction monolayer device with an *n-n* heterojunction and MSM photoconductors to show that the electric field at the *p-n* heterojunction creates the fast photoresponse. We then compare with the *p-n* heterojunction multilayer device to identify factors limiting performance.

Of the monolayer-based devices with comparable channel lengths ($L_{MoS2}$ = 4 µm) and biasing conditions ($V_D$ = -8 V), the *p-n* heterojunction device has the fastest risetime (Fig. 4a). The large electric field in the depletion region of the *p-n* heterojunction (> 3 × 10$^5$ V/cm) rapidly sweeps out photogenerated carriers and minimizes trapping in the active region. The enhanced exciton dissociation[43] also increases the photocurrent. The slower rise time and fall time of $n^+$-$n$ silicon/MoS$_2$

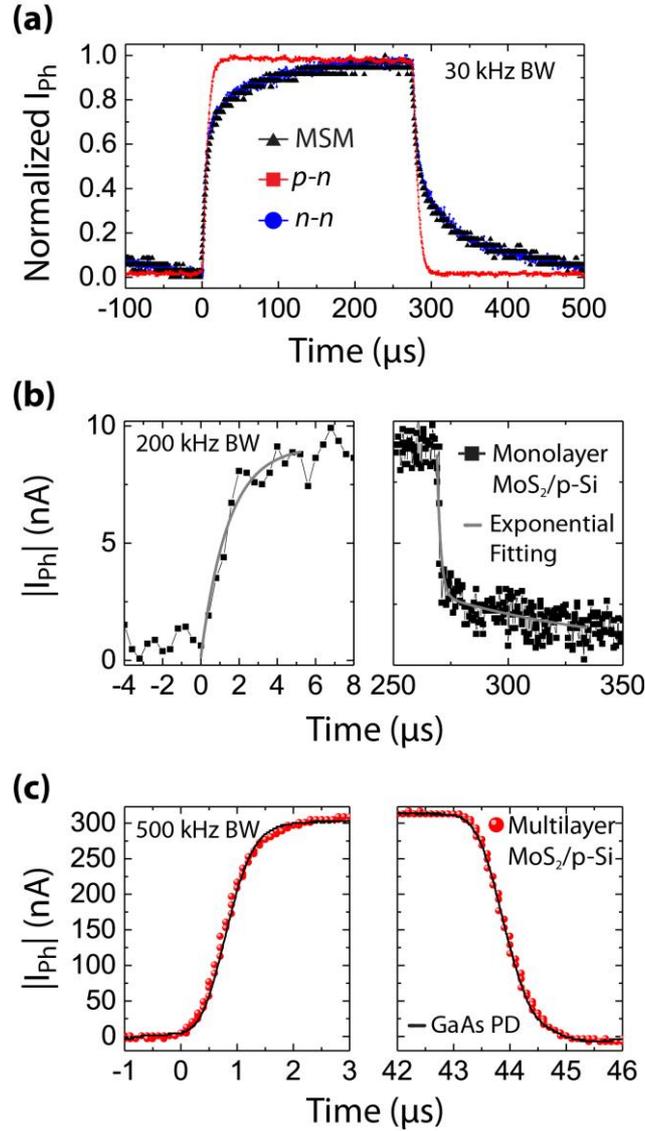

**Figure 4.** (a) Transient photocurrent for *p-n* heterojunction (red squares), *n-n* heterojunction (blue circles) and MSM devices (black triangles) acquired with a temporal resolution of 12 μs (30 kHz bandwidth) at $V_D = -8$ V. (b) Transient photocurrent of monolayer $MoS_2$/Si nanowire *p-n* heterojunction (black squares) measured with a high-speed current preamplifier (200 KHz bandwidth) at $V_D = -8$ V. Rise and fall times are fit with single and bi-exponentials (grey lines), and equal $t_{rise} = 1.4$ μs and $t_{fall} = 1.6$ μs (80 μs), respectively. (c) Transient photocurrent of a multilayer *n*-$MoS_2$/*p*-Si nanowire device (red circles) acquired with a 500 kHz bandwidth preamplifier at $V_D = -6$ V. The photoresponse of a GaAs photodetector is plotted at same time-scale with normalized current magnitude relative to the *n*-$MoS_2$/*p*-Si current. The photoresponse times are limited by the temporal resolution of isotype heterojunction and MoS$_2$ MSM devices ($t_{\text{rise/fall}}$ = 110 µs) reveal the stronger influence of carrier trapping and detrapping in what is predominantly a photoconductive response, as these devices have much smaller electric fields. The maximum electric field in the channel of the photoconductive MoS$_2$ devices, assuming zero contact resistance, is $1.25 \times 10^4$ V/cm, and the simulated field at the heterojunction of the *n-n* device is $1 \times 10^5$ V/cm (Fig. S4). We note that a non-zero contact resistance would further decrease the electric field in the channel of the MSM device and reduce the photocurrent for the case of symmetric contacts.[36] The fastest response time reported for a MoS$_2$-based MSM photodetector to date ($t_{\text{rise}}$ = 2 µs and $t_{\text{fall}}$ = 115 µs) was achieved with chemically bonded molybdenum contacts with small Schottky barrier height ($\Phi_B$ = 0.07 eV).[50]

The higher photocurrent in the *p-n* heterojunction device enables higher bandwidth amplification and the measurement of rise and fall times with higher temporal resolution (Fig. 4b) at the same bias ($V_D$ = -8 V). A single exponential fit to the photocurrent rise gives $t_{\text{rise}}$ = 1.4 µs, and a bi-exponential fit to the decay gives a fast (slow) fall time of 1.6 µs (80 µs). Rise/fall times for previously reported planar Si/MoS$_2$ heterojunctions are slower (10 µs/19 µs[2], 30.5 µs/71.6 µs[1]), as are those of photodetectors with vertically oriented MoS$_2$ on planar Si ($t_{\text{rise}}$ = 3 µs, $t_{\text{fall}}$ = 40 µs[51]). Compared with *planar* Si/MoS$_2$ devices, the Si nanowire maintains a high mobility[52] channel while reducing capacitive coupling to the conductive substrate, resulting in smaller RC time constant, $\tau_{\text{RC}}$.

For all monolayer devices, a relatively small portion of the photoresponse evolves on timescales from hundreds of microseconds to seconds, (Figs. 4ab and S5) likely due to longer-lived interface and surface trap states[50, 53] influenced by the substrate oxide and alumina encapsulation layer.[54] We also note that defects and long-lived charge trapping processes in adsorbates may manifest in photogating, in which photoexcited carriers are trapped and de-trapped in localized states to alter the channel carrier density via field-effect.[53] The time scale of these processes depends on occupation of trap energy levels, which can be varied with the back gate voltage but are generally of the order of milliseconds to seconds.[54] Finally, photothermal contributions that may be present would reflect the timescales of heating and cooling.[55] A small portion of the measured photocurrent might be due to photothermoelectric-effect,[37, 56] previously observed on a millisecond to second time scale in $MoS_2$ based photodetectors.

To analyze the factors influencing the performance of the *p-n* photodetector, we analyzed *p*-Si nanowire/multilayer $MoS_2$ devices that produce a ~30 times larger photocurrent than monolayer devices, enabling higher bandwidth measurements with reduced influence from interfaces. The photocurrent is higher in the multilayer device (Fig. 4c) at comparable bias ($V_D$ = -6 V) due to increased light absorption as well as reduced radiative recombination (due to the indirect band gap of multilayer $MoS_2$) and smaller exciton binding energy[57]. The rise and fall times $t_{rise} = t_{fall} \leq 1$ µs (Figs. 4c and S4) are faster than those of the 1-L $MoS_2$/nanowire device (Fig. 4b), and the fast decay comprises a larger fraction of the photoresponse. We measured the same response time in an ultrafast GaAs *p-i-n* photodetector (Fig. 4c, black line), indicating that the 500 kHz bandwidth of the preamplifier is determining the measured rise time. The inner layers in multilayer $MoS_2$ have fewer structural defects (e.g., surface vacancies) compared with the exposed outer $MoS_2$ layers and compared with monolayer $MoS_2$.[41] The inner layers are also protected from the surface and farther away from the $SiO_2$/$MoS_2$ interface, and thus less affected by interface and surface charges (and trap centers). Thus, it is likely that a reduction in trapping through both shallow and deeper levels improves the performance of the multilayer device beyond changes due to the differences in CVD and exfoliated materials. Interface and surface traps (adsorbates) have been reported to limit the photoresponse time in lateral $MoS_2$ phototransistors.[53, 54 58]

**CONCLUSIONS** In summary, finite element simulations in conjunction with SPCM reveal the formation of a photoresponsive depletion region in a non-planar mixed-dimensional *p*-Si nanowire/*n*-$MoS_2$ heterojunction. The lateral depletion region in $MoS_2$ increases with the reverse bias and explains the spatial distribution of the photocurrent. In contrast, photocurrent in *n*-Si nanowire/$MoS_2$ devices is restricted to the heterojunction. Photoresponse times of less than $\leq 1$ μs were measured for multilayer $MoS_2$/*p*-Si nanowire devices, which is faster than that of monolayer $MoS_2$/*p*-Si nanowire devices studied here and an order of magnitude faster than the reported response times for planar Si/$MoS_2$ heterojunction devices[1, 2] whose response times are likely RC-limited due to the larger channel area. The photoresponse time could be further decreased by directly contacting the $MoS_2$/nanowire heterojunction area and thus minimizing the transit time through $MoS_2$. This work explains how the electrostatics of a non-planar heterojunction device affects charge carrier transport, which is essential for the design of future optoelectronic devices.

MATERIALS AND METHODS

**Materials growth.** P-type Si nanowires were grown in a hot-wall chemical vapor deposition system via the vapor-liquid-solid mechanism utilizing Au nanoparticles of 100 nm diameter. The growth was carried out at 460 °C at a base pressure of 40 Torr with $SiH_4$, $B_2H_6$ (100 ppm in He) and $H_2$ flow rates of 3, 15, and 60 sccm, respectively. The n-type nanowires were grown at 460 °C at a base pressure of 40 Torr with $SiH_4$,

PH$_3$ (200 ppm in He) and H$_2$ flow rates of 3, 30 and 60 sccm, respectively. Monolayer MoS$_2$ was grown by chemical vapor deposition on diced 300 nm thick SiO$_2$/Si wafers (SQI Inc.) as described in detail in our previous report.[59]

**Transfer of nanowires and MoS$_2$.** Nanowires were suspended in solution by sonicating (70 V, 30 s) the growth substrate submerged in isopropyl alcohol (IPA). The nanowires were then drop-cast onto Si/SiO$_2$ substrates placed on a hot plate at 80 °C.

Transfer of 1-L MoS$_2$ on Si/SiO$_2$ substrates with nanowires involved the following steps: (1) poly(methylmethacrylate) (PMMA) A4 950 (Microchem Corp.) was spin-coated (3000 rpm, 60 s) on the MoS$_2$/SiO$_2$/Si substrates and subsequently annealed in air (180 °C, 10 min). (2) The sample was immersed in 3M potassium hydroxide (KOH) solution for ~24 hours to etch away the SiO$_2$ and release the PMMA/MoS$_2$ thin film from the Si/SiO$_2$ substrate. (3) A clean glass slide was used to transfer and float the delaminated PMMA/MoS$_2$ thin film in a clean DI water bath. Next, the film was transferred on to the target substrate (Si/SiO$_2$ chip with nanowires) and annealed for 5 min at 80 °C and 15 min at 150 °C. (4) The sample was then immersed in chloroform (HPLC Plus Sigma) for 24 h to dissolve the PMMA, and finally rinsed in IPA and blow-dried with nitrogen.[59] To remove PMMA residues, the devices were annealed at 250 °C for 2 hours in an H$_2$/Ar atmosphere in a tube furnace. To realize multilayer MoS$_2$/nanowire heterojunctions, MoS$_2$ was mechanically exfoliated on substrates with nanowires. Prior to MoS$_2$ the substrate was dipped for 3 s in a buffered oxide etch (BOE) 1:5 (HF:NH$_4$F) to remove the native SiO$_2$.

**Device fabrication.** To realize nanowire/MoS$_2$ heterojunction devices, contacts were fabricated in a lift-off process, which involved the following steps: (1) PMMA A4 950 (MicroChem Corp.) was spin-coated at 4000 rpm for 45 s followed by baking on a hotplate at 180 °C for 90 s. (2) Contacts were patterned via electron-beam lithography (EBL) using a FEI Quanta environmental SEM; operated at a working distance of 6.6 mm and an accelerating voltage of 30 kV, a beam current of 500 pA, and dosages of 330-350 µC/cm$^2$. (3) The substrate was developed for 75 s in methyl isobutyl ketone (MIBK):IPA (1:3) solution and 15 s in MIBK:IPA (1:1) solution. (4) The Si native SiO$_2$ was etched off in BOE 1:5 (HF:NH$_4$F) for 5 s. Subsequently, titanium (3 nm) and gold (50 nm) were deposited at a pressure below $2 \times 10^{-7}$ Torr using an electron-beam evaporator (Mantis deposition Inc.). (5) Lift-off was done in N-methyl-2-pyrrolidone (NMP) at 75 °C for 1 hour. Finally, the devices were rinsed in IPA and blow-dried with nitrogen. Monolayer MoS$_2$ heterojunction devices were encapsulated with a 10 nm thick atomic layer deposition (ALD) grown Al$_2$O$_3$ layer.

**Scanning photocurrent microscopy.** SPCM was performed using a confocal microscope (WiTec, Alpha 300R) in ambient. A coherent white light source (NKT Photonics) was tuned to a wavelength of 500 nm and a power of 20 µW, and modulated with an acousto-optic tunable filter (AOTF) modulator (NKT Photonics) at 1837 Hz. The laser was fiber-coupled into the microscope optics and focused ($1 \times 1$ µm$^2$) onto the sample with a 50× objective. We estimate a fluence of 32 µJ/cm$^2$ using the above parameters together with the laser repetition rate of 78 MHz. Source and drain electrodes were contacted with radio frequency (RF) microprobes (Quater Research and Development, 20340) terminated with SMA (SubMiniature version A, SMA) connectors, and the photocurrent was transmitted via coaxial RF cables. The source electrode (MoS$_2$) and the global back gate (Si substrate) were grounded while the applied bias to the drain electrode (Si nanowire) was varied with the internal power supply of the lock-in amplifier (Signal recovery, 7265). The photocurrent was amplified using a variable gain current preamplifier (DL Instruments, 1211, 60 KHz BW) and detected with a lock-in amplifier (Signal recovery, 7265) at 20 mV sensitivity and a time constant of 5 ms. Photocurrent images were acquired by scanning the sample at a 50 nm resolution using a piezo-controlled stage, and at an integration time of 15 ms at each measurement point.

**Time-resolved photocurrent measurements.** Time-resolved photocurrent measurements were conducted using the above described SPCM setup. However, custom-made dc battery voltage sources replaced the ac power supply to reduce noise. The generated photocurrent was amplified using one of the following amplifiers: a low-noise amplifier (DL instruments, 1211, 50 KHz bandwidth), a variable gain preamplifier (Femto, DLPCA-200, 500 KHz bandwidth), or a high-speed preamplifier (Femto, HCA-40M-100K-C, 40 MHz bandwidth). The light was modulated either at 1837 Hz, 10837 Hz or 33837 Hz, and the signal was sampled by an oscilloscope (Tektronix, TDS 2014, 100 MHz BW) that was synced with the AOTF modulation. The signal was averaged over 128 acquisitions to cancel out noise; the successive traces are identical, and the averaged waveform is an accurate rendition of the signal at the oscilloscope input. The system response was measured using an unmounted GaAs p-i-n photodetector (Cosemi, SPD2014, 12 GHz BW) contacted with microprobes.

**Atomic force microscopy.** Amplitude modulation ("tapping" mode) AFM was carried out with a Dimension Icon AFM (Bruker Inc.) in ambient. AFM probes with a backside reflective aluminum coating and a silicon tip with a nominal radius of 10 nm (TESPA, Bruker Inc.) were used. The topography was measured by maintaining the amplitude of the first cantilever resonance ($f_{1st} \cong 320$ kHz) at a predefined amplitude set point of approximately 13 nm corresponding to 65% of the free vibrational amplitude (20 nm).

ASSOCIATED CONTENT

**Supporting Information.**

The supporting information is available free of charge on the ACS publications website. Additional details on experimental methods, electrical and optical characterization, simulation data, scanning photocurrent microscopy and time-resolved photocurrent data, as well as scanning electron microscopy images (PDF).

AUTHOR INFORMATION

**Corresponding Author**

*Lincoln J. Lauhon, E-mail: lauhon@northwestern.edu

**Author Contributions**

All authors have approved the final version of the manuscript.

**Funding Sources**

This research was supported by the 2-DARE program (NSF EFRI-1433510) and the Materials Research Science and Engineering Center (MRSEC) of Northwestern University (NSF DMR-1720139). CVD growth of $MoS_2$ was supported by the National Institute of Standards and Technology (NIST CHiMaD 70NANB14H012).

**Notes**

The authors declare no competing financial interest.

ACKNOWLEDGMENT A.H. acknowledges the support of a Research Fellowship from the Deutsche Forschungsgemeinschaft (Grant HE 7999/1-1). H.B acknowledges support from the NSERC Postgraduate Scholarship-Doctoral Program and the National Science Foundation Graduate Research Fellowship. This work made use of the Northwestern University NUANCE Center and the Northwestern University Micro/Nano Fabrication Facility (NUFAB), which are partially supported by the Soft

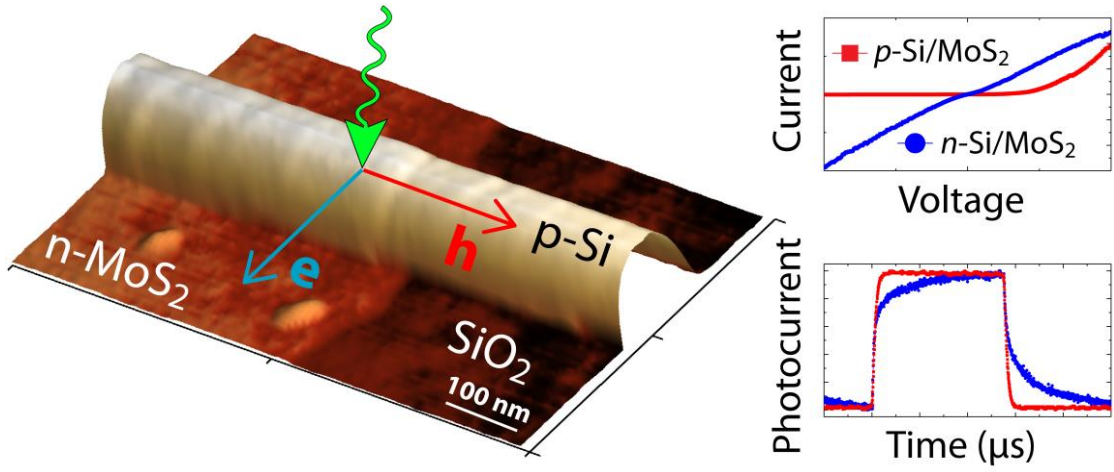